\documentclass{article}

\usepackage{arxiv}

\usepackage[utf8]{inputenc} 
\usepackage[T1]{fontenc}    
\usepackage{hyperref}       
\usepackage{url}            
\usepackage{booktabs}       
\usepackage{amsfonts}       
\usepackage{nicefrac}       
\usepackage{microtype}      
\usepackage{graphicx}
\usepackage[numbers]{natbib}
\usepackage{doi}
\usepackage{enumitem}
\usepackage{makecell}
\usepackage{titlesec}

\setlist{leftmargin=3.0mm}

\newcommand{\overbar}[1]{\mkern 1.5mu\overline{\mkern-1.5mu#1\mkern-1.5mu}\mkern 1.5mu}
\titlespacing\subsection{0pt}{12pt plus 4pt minus 2pt}{0pt plus 4pt minus 2pt}

\title{Toward Accurate Interpretable Predictions of Materials Properties within Transformer Language Models}

\author{ \href{https://orcid.org/0000-0001-6117-5662}{\includegraphics[scale=0.07]{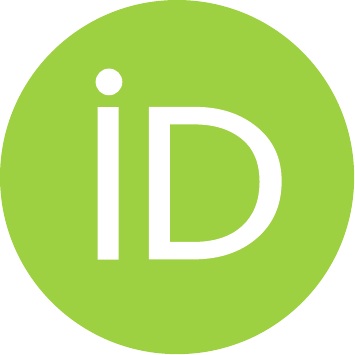}\hspace{1mm}Vadim Korolev}\thanks{\textit{Email address}: \texttt{korolev@colloid.chem.msu.ru}} \\
	Department of Chemistry\\
	Lomonosov Moscow State University\\
	Moscow 119991, Russia\\
	\And
	\href{https://orcid.org/0000-0002-1503-3679}{\includegraphics[scale=0.07]{orcid.pdf}\hspace{1mm}Pavel Protsenko} \\
	Department of Chemistry\\
	Lomonosov Moscow State University\\
	Moscow 119991, Russia\\
}

\renewcommand{\shorttitle}{Toward Accurate Interpretable Predictions of Materials Properties}

\begin{document}
\maketitle

\begin{abstract}
Property prediction accuracy has long been a key parameter of machine learning in materials informatics. Accordingly, advanced models showing state-of-the-art performance turn into highly parameterized black boxes missing interpretability. Here, we present an elegant way to make their reasoning transparent. Human-readable text-based descriptions automatically generated within a suite of open-source tools are proposed as materials representation. Transformer language models pretrained on 2 million peer-reviewed articles take as input well-known terms, e.g., chemical composition, crystal symmetry, and site geometry. Our approach outperforms crystal graph networks by classifying four out of five analyzed properties if one considers all available reference data. Moreover, fine-tuned text-based models show high accuracy in the ultra-small data limit. Explanations of their internal machinery are produced using local interpretability techniques and are faithful and consistent with domain expert rationales. This language-centric framework makes accurate property predictions accessible to people without artificial-intelligence expertise.
\end{abstract}

\keywords{property prediction \and explainable artificial intelligence \and language models \and transformers \and fine-tuning}

\section{Introduction}
\label{sec:introduction}
Artificial intelligence (AI) is increasingly perceived as the fourth pillar of modern science\cite{von2020introducing} rather than a tool complementary to the previous three, namely experiment, theory, and simulation. In the materials science realm, a data-driven approach has been successfully employed to capture complex structure–property relationships. In particular, AI techniques have pushed the frontiers of high-throughput computational screening\cite{meredig2014combinatorial,faber2016machine,seko2017representation,bartel2018physical}, inverse materials design\cite{noh2019inverse,korolev2020machine,yao2021inverse,ren2022invertible}, interatomic potential development\cite{chen2022universal,sauceda2022bigdml,choudhary2023unified}, and crystal structure prediction\cite{ryan2018crystal,podryabinkin2019accelerating,liang2020cryspnet}. Users demand that supervised machine learning (ML) models involved in the above tasks be accurate first. Driven by this demand, AI practitioners among materials scientists have developed increasingly complex models by elaborating data representations. A retrospective view of the evolution of graph neural networks in the field\cite{reiser2022graph} should serve as a definitive example: the models that hold global state attributes\cite{chen2019graph} and many-body interactions\cite{choudhary2021atomistic} have shown growing improvement in performance relative to neural networks trained on a compact set of node and edge features\cite{xie2018crystal}. The state-of-the-art architectures aimed at materials property predictions may contain many thousands or millions of trainable weights (in other tasks, billions\cite{brown2020language} and trillions\cite{fedus2021switch}) making a good understanding of internal machinery intractable for human beings. Unsurprisingly, the explainability problem gives rise to distrust and hinders wider applications of ML algorithms. It should be noted that there are special scenarios where the use of uninterpretable black boxes may be appropriate\cite{holm2019defense}. Nevertheless, the explanation of model reasoning (“why is it the answer?”) is a desideratum of effective AI in general\cite{doshi2017towards} and specific\cite{meredig2019five} contexts.

Explainable AI\cite{gunning2019xai} (XAI) is an umbrella term for algorithms intended to make their decisions transparent by providing human-understandable explanations. Following the proposed taxonomies\cite{adadi2018peeking,arrieta2020explainable,das2020opportunities}, one can differentiate XAI methods based on multifaceted but nonorthogonal dualities: model-specific vs model-agnostic, intrinsic vs \emph{post hoc}, and local vs global explainers. Despite the impressive diversity, only a few XAI approaches are applied in materials science actively. The most remarkable techniques are examined below; for more details, please refer to recent reviews\cite{esterhuizen2022interpretable,zhong2022explainable,oviedo2022interpretable}. First, we would like to highlight supervised ML algorithms that have inherent transparency. Linear regression models and their extension, generalized additive models\cite{lou2012intelligible,lou2013accurate}, provide weight coefficients as an importance metric of relevant features\cite{xin2012predictive,allen2022machine,esterhuizen2020theory}. Decision trees\cite{breiman2017classification} are another method that has shown off-the-shelf transparency. For example, probability values in terminal nodes of classification models reveal the combinations of splitting criteria leading to preferable output\cite{carrete2014nanograined,fernandez2016geometrical,wu2020machine}. The next approach involves the derivation of analytical expressions of structure–property relationships. Symbol regression\cite{wang2019symbolic} and compressed sensing methods such as LASSO\cite{tibshirani1996regression} and SISSO\cite{ouyang2018sisso} make it possible to access sets of solutions competitive in terms of accuracy and complexity\cite{ghiringhelli2015big,andersen2019beyond,bartel2019new,weng2020simple,singstock2021machine}. Finally, there are \emph{post hoc} local explainers that quantify feature importance levels when analyzing opaque ML models. Among such XAI schemes, the SHapley Additive exPlanations\cite{lundberg2017unified} (SHAP) suite seems to be dominant in materials science applications\cite{liang2019phillips,korolev2020transferable,jablonka2021data,georgescu2021database,zhang2021predicting,marchenko2021relationships,korolev2021parametrization,anker2022extracting,lu2022fly}.

The common thread in most of the explainability-aware studies mentioned above relates to employing low-dimensional handcrafted features as input to a training model. Besides the notorious tradeoff between accuracy and explainability of ML algorithms\cite{lipton2018mythos}, a similar compromise is seen in materials representations. Going beyond simplistic physicochemical descriptors, more advanced featurization (for instance, physics-inspired\cite{musil2021physics}) schemes can hardly be interpreted in terms familiar to domain specialists. Nonetheless, there is an alternative to tabular-like representations; as a consequence, other XAI techniques may come into play. Graph neural networks rooted in deep geometric learning\cite{battaglia2018relational,bronstein2021geometric} successfully cope with data irregularities. In particular, periodic atomistic systems can be processed in a natural way if the concept of a crystal graph\cite{chen2019graph,choudhary2021atomistic,xie2018crystal,fedorov2017crystal,korolev2019graph,park2020developing,cheng2021geometric,omee2022scalable} is introduced. The propagation and aggregation of information contained in node and edge attributes via message passing and pooling operators are aimed at describing pair-wise and higher-order interactions. To date, graph neural networks have archived state-of-the-art performance in predicting a plethora of structure–property relationships\cite{reiser2022graph,fung2021benchmarking}. The uncovering of such black box models in order to gain chemical insights is addressed by the development of graph-specific XAI techniques\cite{ying2019gnnexplainer,noutahi2019towards,luo2020parameterized} and by the adaptation of methods from other domains\cite{pope2019explainability,sun2021explanation,huang2022graphlime}. The field is in its infancy, and therefore explainability-aware studies exploring graph neural networks in materials science are sparse and few\cite{raza2020towards,hsu2022efficient,chen2022interpretable}.

There is no community consensus on defining explainability owing to the diversity in XAI approaches and problems being solved. Nevertheless, attempts to specify the term\cite{doshi2017towards,montavon2018methods,gilpin2018explaining,guidotti2018survey} are united by the idea that the perceiver (domain expert) is as important as the explainer (XAI algorithm). Moreover, cognitive abilities of the former limit model understanding\cite{schwartz2022should}, which is the primary reason why researchers are forced to consider simple features in supervised ML if model reasoning is simple in the first place. We stress that natural language representation of materials is an optimal way to archive interpretability by human beings. The corresponding AI field, natural language processing\cite{hirschberg2015advances}, has already found successful applications in materials science such as named entity recognition\cite{weston2019named,gupta2022matscibert,trewartha2022quantifying} and paragraph classification\cite{gupta2022matscibert,huo2019semi,huang2022batterybert}. At the same time, the potential of natural language features in materials property prediction is fully unexplored to the best of our knowledge.

In this study, we present a language-centric framework able to reconcile high accuracy and interpretability of the prediction of materials properties. Attention-based neural networks trained on text descriptions are thoroughly compared with graph neural networks, including compositional and structure-aware architectures. A classic ML algorithm, random forest, built on force-field inspired descriptors is included into the benchmark as well. We demonstrate remarkable scalability of the language models that allow to achieve state-of-the-art performance in a small-data regime. In certain cases, transformers trained on human-readable features surpass graph neural networks despite training-dataset size. The interpretability of our approach is estimated in terms of faithfulness and plausibility. As the analysis showed, XAI approaches can generate sufficient and comprehensible explanations that are consistent with expert decision-making.

\section{Results and Discussion}
\label{sec:results}

\subsection{General workflow}
\begin{figure}[ht]
  \centering
  \includegraphics[width=11cm]{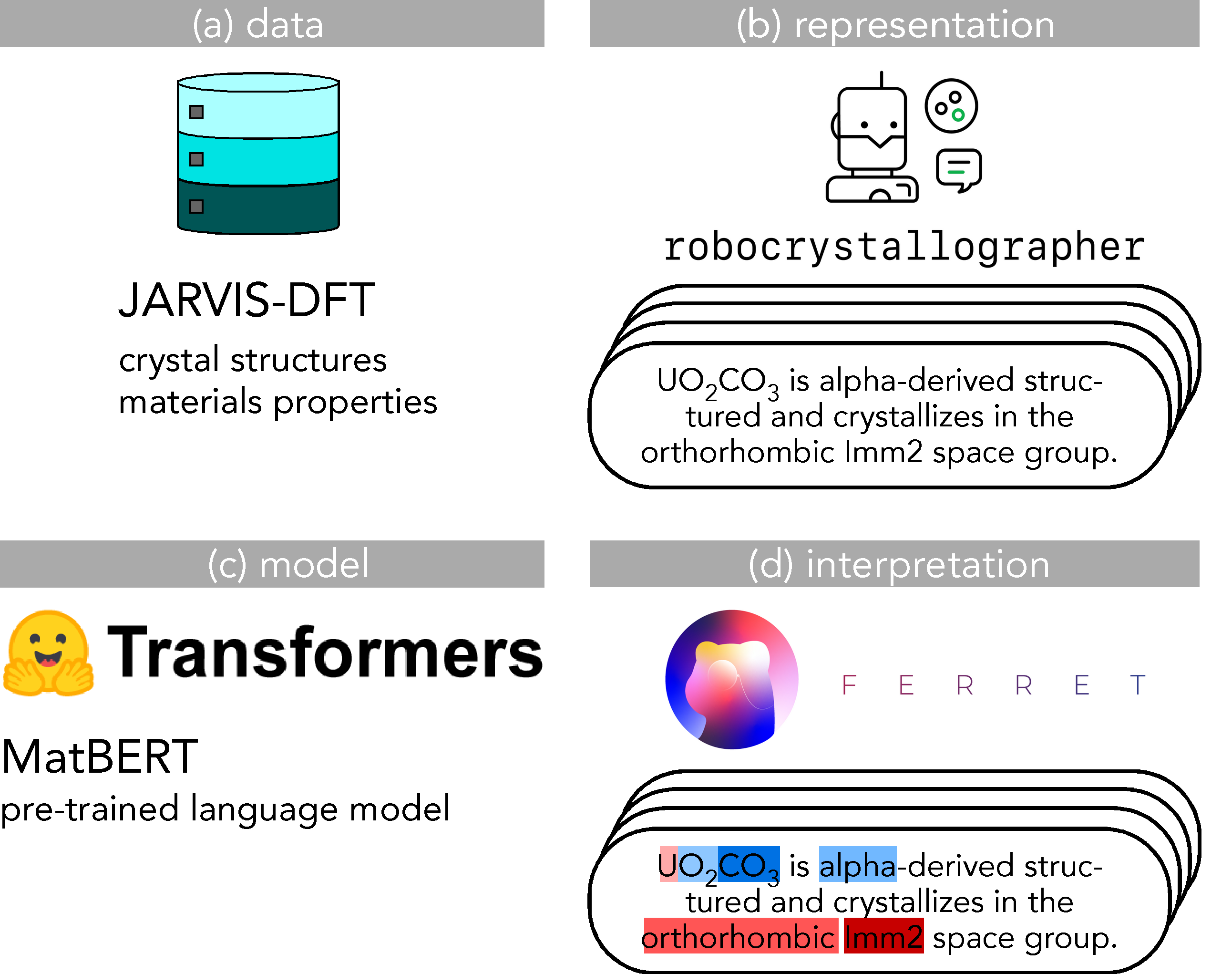}
  \caption{An overview of the language-centric approach. (a) Initial data are taken from an open computational database containing crystal structures and a diverse set of physical properties calculated at the level of density functional theory (DFT). (b) Proceeding from crystal structures, we generate text-based descriptions via an automatic toolkit. Local, semilocal, and global environment features are taken into account. (c) Neural networks capable of handling a natural language are trained on the text-based descriptions to classify materials. (d) \emph{Post hoc} explainability techniques help to rationalize algorithm decisions at the level of tokens.}
  \label{fig:fig1}
\end{figure}

The general workflow of the study is presented in Figure \ref{fig:fig1}. The first three stages are the preparation of the main supervised ML components\cite{haghighatlari2020learning}: an input dataset, feature representation, and a predictive algorithm. We start by considering the crystal structures and corresponding property values taken from the Joint Automated Repository for Various Integrated Simulations\cite{choudhary2020joint} (JARVIS). The following diverse set of endpoints is taken into account: energy above the convex hull, the magnetic moment, a band gap, spectroscopic limited maximum efficiency (SLME), and topological spin-orbit spillage. All the data were originally obtained at the density functional theory (DFT) level in accordance with standardized calculation procedures. High-throughput computational databases\cite{curtarolo2012aflow,saal2013materials,jain2013commentary} such as JARVIS-DFT help an investigator to focus on ML tasks (e.g., feature and algorithm benchmarking and model fine-tuning) and to simplify data preprocessing. At the second step, we for the first time implement human-readable text descriptions as materials representation for supervised ML. We use the Robocrystallographer library\cite{ganose2019robocrystallographer} to automatically generate the proposed representation for thousands of crystal structures, but a similar content written by a human crystallographer is also acceptable. At the third step, advanced language models, namely transformers\cite{trewartha2022quantifying,vaswani2017attention,devlin2018bert}, are utilized to extract structure–property relationships based on text descriptions generated at the previous step. All the prediction tasks are examined in the classification mode. The last step of our workflow is designed to assess interpretability of the language-centric approach. The trained models are processed within the suite of \emph{post hoc} local XAI techniques\cite{attanasio2022ferret}.

\subsection{Model performance}
Aside from the input representation and network architecture, several other factors directly influence language model performance. First, various vocabularies can be applied during tokenization, which is a procedure of splitting text into elementary bits of information. Second, the model can be optionally pretrained on a large text corpus before use in downstream tasks. We present results of training of three models differing in the above characteristics. The transformer models combined with a general and domain-specific tokenizer without pretraining are designated as BERT and BERT-domain, after the name of the base architecture: Bidirectional Encoder Representations from Transformers\cite{devlin2018bert}. The transformer model pretrained on the corpus of materials science papers\cite{trewartha2022quantifying} (MatBERT) and combined with the domain-specific tokenizer is tested as well. We would like to integrate the presented models into the landscape of modern ML models predicting materials properties. To do so, the following algorithms are included into the benchmark for a comparison with the models mentioned above. Random forest\cite{breiman2001random} trained on Classical Force-Field Inspired Descriptors\cite{choudhary2018machine} (RF-CFID) serves as a representative classic ML algorithm built on tabular features. A deep neural network designated as Representation Learning from Stoichiometry\cite{goodall2020predicting} (Roost) typifies advanced models trained on chemical compositions. Finally, Atomistic Line Graph Neural Network\cite{choudhary2021atomistic} (ALIGNN) is examined as a state-of-the-art predictor of materials properties.

As a target metric, we calculate the Matthews correlation coefficient (MCC). The metric is recognized as a reliable statistical measure and is preferable to other binary classification metrics, including accuracy and the F1 score\cite{chicco2020advantages}. MCCs for all the above models and endpoints are provided in Table \ref{tab:tab1}. MatBERT works surprisingly well, manifesting state-of-the-art performance in four cases out of five. ALIGNN has the highest MCC only on magnetic/nonmagnetic classification. The overall MCC across all endpoints equals 0.74 and 0.72 for MatBERT and ALIGNN, respectively. RF-CFID and Roost show worse performance than ALIGNN, with one exception. Roost has the second-highest MCC on energy above the convex hull. The same trends are observed for accuracy (Table S1) and the F1 score (Table S2). The relatively high efficiency of the structure-agnostic model (Roost) contradicts previous results. Bartel et al. have demonstrated significant improvement in stability predictions owing to inclusion of crystal structure in representation\cite{bartel2020critical}. On the other hand, it is important to note that another endpoint (decomposition energy) has been addressed by those researchers. The interplay of model architecture and the thermodynamic stability criterion seems to be a promising avenue of future work and is beyond the scope of this study.

\begin{table}[ht]
\centering
\caption{\label{tab:Table1}Model performance in terms of the Matthews correlation coefficient (MCC). The best coefficient for each endpoint is \textbf{boldfaced}; the second-best one is \underline{underlined}.}
\setlength{\tabcolsep}{0.5em}
\renewcommand{\arraystretch}{1.4}
\begin{tabular}{ l c c c c c }
 \hline
 & \makecell{energy above \\ the hull}	& \makecell{magnetic \\ moment} & band gap & SLME & \makecell{spin-orbit \\ spillage}\\
 \hline
 RF-CFID & 0.791 ± 0.012	& 0.735 ± 0.012 & 0.800 ± 0.013 & 0.595 ± 0.018 & 0.492 ± 0.027\\
 Roost & \underline{0.885 ± 0.005} & 0.762 ± 0.009 & 0.794 ± 0.020 & 0.580 ± 0.019 & 0.482 ± 0.025\\
 ALIGNN & 0.878 ± 0.010 & \textbf{0.793 ± 0.009} & \underline{0.827 ± 0.011} & \underline{0.615 ± 0.027} & \underline{0.507 ± 0.026}\\
 BERT & 0.788 ± 0.011 & 0.674 ± 0.014 & 0.747 ± 0.014 & 0.446 ± 0.026 & 0.401 ± 0.027\\
 BERT-domain & 0.841 ± 0.013 & 0.727 ± 0.011 & 0.791 ± 0.011 & 0.52 ± 0.04 & 0.464 ± 0.026\\
 MatBERT & \textbf{0.901 ± 0.005} & \underline{0.788 ± 0.007} & \textbf{0.845 ± 0.011} & \textbf{0.629 ± 0.017} & \textbf{0.519 ± 0.022}\\
\end{tabular}
\label{tab:tab1}
\end{table}

We can speculate that the superiority of the presented language-centric approach over others rests on knowledge enrichment rather than the model architecture. For instance, MatBERT significantly outperforms BERT-domain (an overall MCC of 0.67), whereas the only difference between them is pretraining of the former. In particular, BERT-domain was trained on 1.6–10.1 million tokens depending on an endpoint. MatBERT was trained on 8.8 billion\cite{trewartha2022quantifying} + 1.6–10.1 million tokens, taking into account the masked language modeling of the original model. Therefore, the effective dataset size increases by approximately three to four orders of magnitude via fine-tuning.

The approach applied herein—fine-tuning—belongs to the transfer learning paradigm. In materials science, models are usually pretrained on labeled data in a supervised manner\cite{yamada2019predicting,jha2019enhancing}. By contrast, language models, such as transformers, provide a great opportunity to capture domain knowledge within self-supervised learning. The above-mentioned masked language modeling is a vivid representative of such techniques. The proposed language-centric approach allows to profitably incorporate a massive source of scientific knowledge (journal publications) into a workflow intended for materials property predictions. Henceforth, domain-specific text corpora should be seen as an alternative to high-throughput DFT databases\cite{choudhary2020joint,curtarolo2012aflow,saal2013materials,jain2013commentary} containing thousands of crystal structures and calculated properties in the context of transfer learning. Moreover, predictors based on a text description can benefit from processing of large samples of papers even without the training of language models on them. BERT-domain clearly surpasses BERT in terms of the MCC, accuracy, and F1 score (Tables 1, S1, and S2) for all endpoints, exclusively due to the domain-specific tokenizer trained on 2 million materials science articles. This result proves the importance of using a domain-specific vocabulary because the general-purpose BERT tokenizer is unable to meaningfully process chemical formulas, space group symbols, and other data. Another remarkable feature of such preprocessing is that vocabulary construction, even on large corpora, takes negligible computing time. By analogy with corpora of academic texts and computational databases, we view domain-specific tokenization as a low-cost alternative to fine-tuning within a general-purpose vocabulary.

\begin{figure}[ht]
  \centering
  \includegraphics[width=14cm]{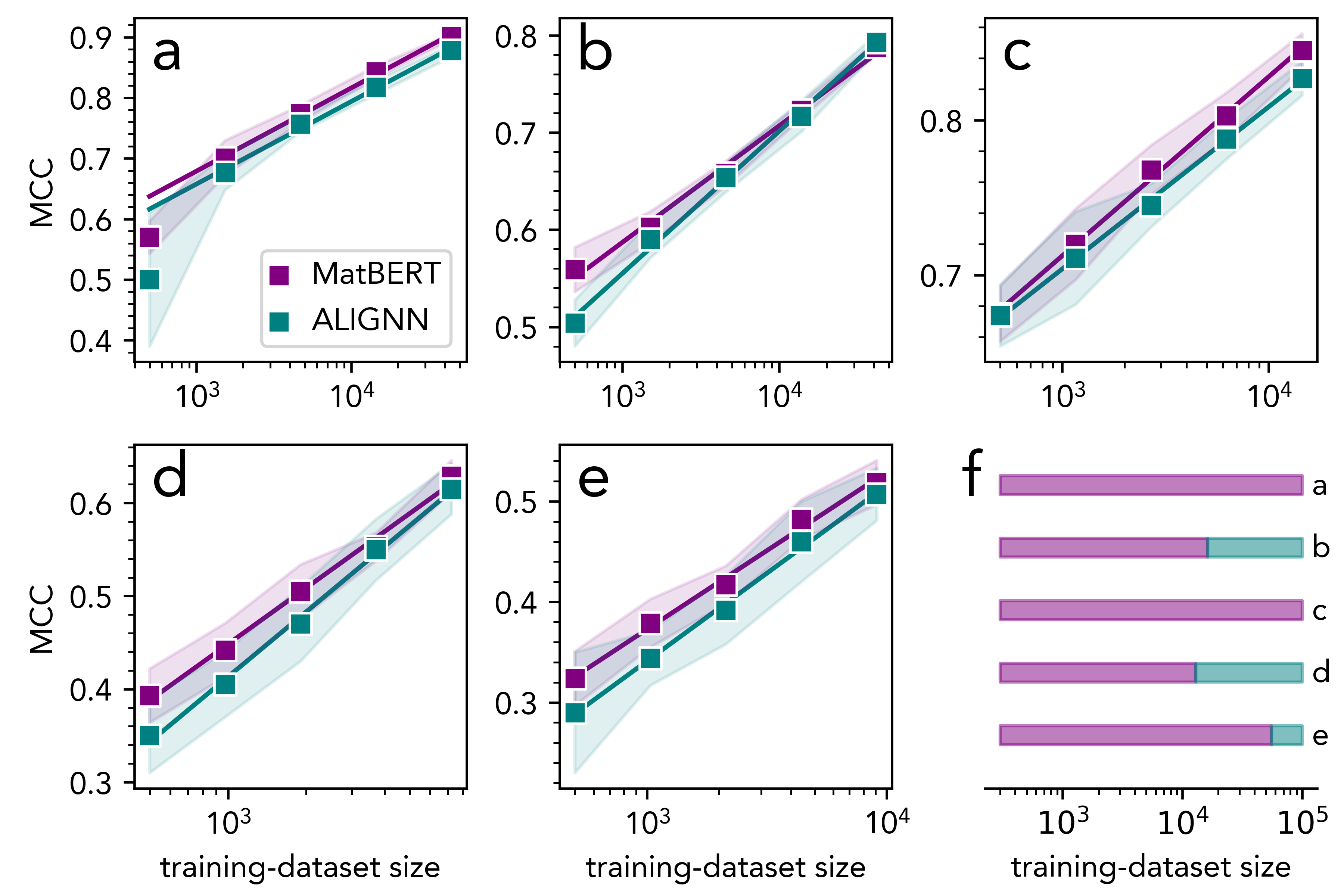}
  \caption{Scalability of Atomistic Line Graph Neural Network (ALIGNN) and Bidirectional Encoder Representations from Transformers model pretrained on Materials science literature (MatBERT). Matthews correlation coefficients (MCCs) are shown as a function of training-dataset size for the following classification tasks: (a) energy above the convex hull, (b) the magnetic moment, (c) a band gap, (d) spectroscopic limited maximum efficiency (SLME), and (e) topological spin-orbit spillage. The solid lines denote linear fitting of the data; the smallest dataset (499 entities) in the energy-above-the-convex-hull task is not included into the analysis because of a severe deviation from linear behavior. The shaded areas are standard deviation across 10-fold cross-validation. (f) Regions of dominance of two examined models in terms of the MCC are marked by the corresponding color.}
  \label{fig:fig2}
\end{figure}

It is well known that ML model performance depends on training-dataset size\cite{zhang2018strategy}. Moreover, predictive algorithms significantly differ in sensitivity to the growth of available data\cite{himanen2020dscribe,langer2022representations}. To clarify this issue for the considered endpoints, we retrain two best-performing models, MatBERT and ALIGNN, on a part of the original training datasets. Figure \ref{fig:fig2} shows a strong linear dependency (${R}^{2}$ > 0.99) for a logarithmic scale. MatBERT is more accurate in the ultra-small data limit (has a systematically higher intercept value in the linear equations), in good agreement with the hypothesis that fine-tuning affords an increase in effective dataset size. Assuming that the linear approximation holds true as the training-dataset size grows, we identify limits of superiority of MatBERT over ALIGNN (Figure \ref{fig:fig2}f). ALIGNN has the highest MCC with a training-dataset size exceeding 16297, 12974, and 55533 entities in the case of the magnetic moment, SLME, and spillage, respectively. On the contrary, MatBERT dominates in classification of energy above the convex hull and a band gap regardless of data availability (has simultaneously higher intercept and slope values in the linear equations). This is a surprising result because generally, higher scalability of a graph neural network is expected for the following reason. The ALIGNN architecture incorporates information-rich structure representation, which is capable of absorbing subtle crystallographic features in contrast to human-readable features implemented in MatBERT. Consequently, each additional training point should potentially enrich the graph neural network more than the language model. Regardless of whether the state-of-the-art performance of the presented approach is confirmed on larger training datasets, MatBERT shows unexpectedly good performance at the scale of thousands of training samples.

\subsection{Model explanation}
Now we are going to evaluate interpretability of the language-centric approach. According to the prominent framework introduced in ref.\cite{deyoung2019eraser}, two peculiar aspects are taken into account. First, we examine the ability of XAI techniques to correctly reflect internal machinery of a predictor, i.e., faithfulness. Second, the consistence of explainer output within human reasoning, aka plausibility, is evaluated. We would like to emphasize that the proposed materials representation (entity level input into a predictor) is a human-readable text description separated into a sequence of tokens. To explain decisions of the black box model at the level of distinct entities, local \emph{post hoc} XAI techniques are employed. The choice of feature importance methods is explained by the resemblance between how such approaches represent ML reasoning and how human beings tend to perceive a natural language, by highlighting the most meaningful parts of a text. Four techniques are implemented to identify tokens most impactful on the prediction: saliency map extraction via computation of input gradients\cite{simonyan2013deep} (hereinafter referred to as SM), integrated gradients\cite{sundararajan2017axiomatic} (IG), Local Interpretable Model-agnostic Explanations\cite{ribeiro2016should} (LIME), and SHapley Additive exPlanation\cite{lundberg2017unified} (SHAP). We apply several explainers to achieve suboptimal results for a specific task because there is no \emph{a priori} knowledge about which explainer shows state-of-the-art performance.

\begin{figure}[ht]
  \centering
  \includegraphics[width=14cm]{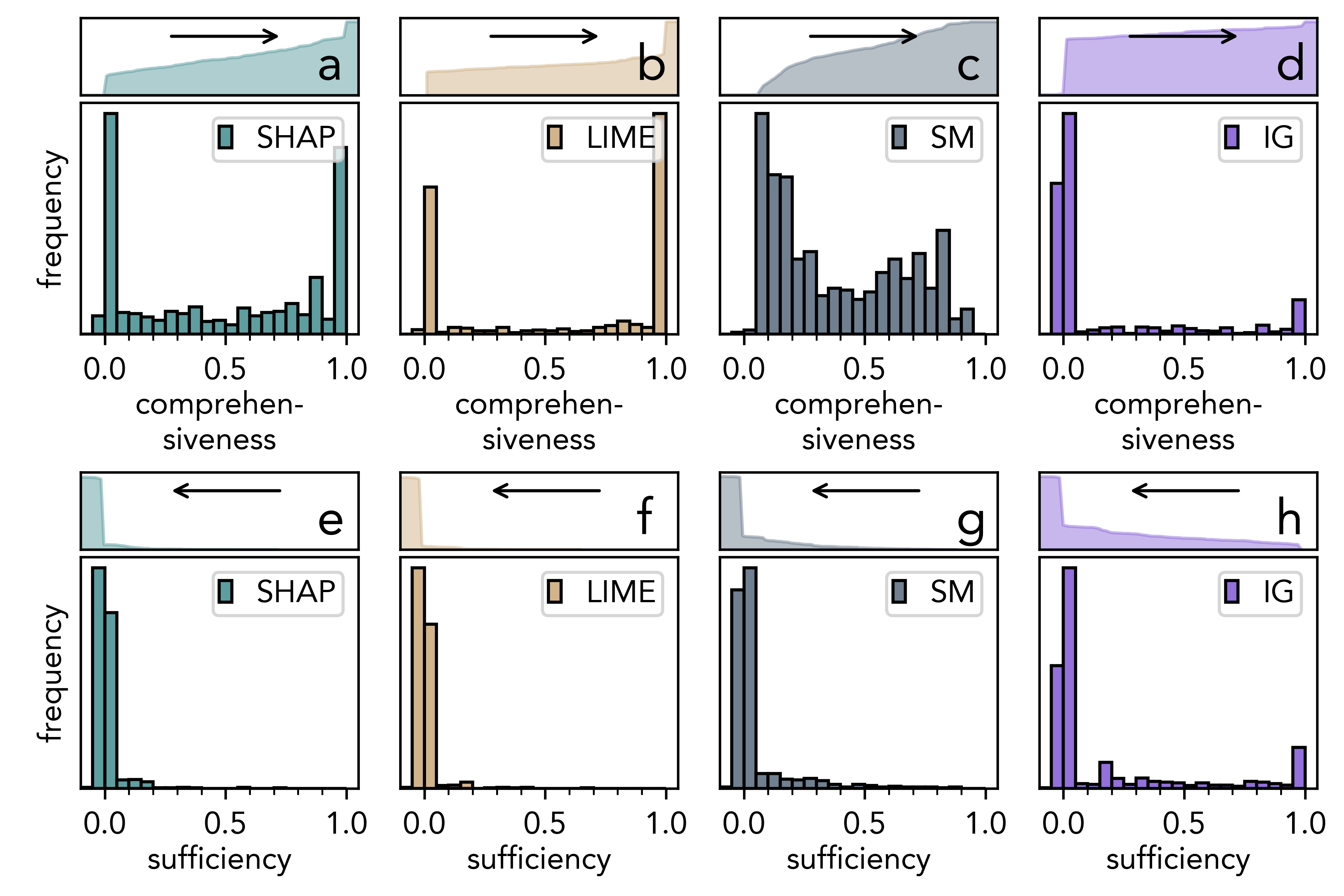}
  \caption{Faithfulness metrics determined within \emph{post hoc} local explainability techniques. The MatBERT model for band gap classification is examined. Each subplot contains a distribution of calculated metric values (comprehensiveness or sufficiency) and a respective cumulative curve. The results are presented for the following explainers: (a, e) SHapley Additive exPlanations (SHAP), (b, f) Local Interpretable Model-agnostic Explanations (LIME), (c, g) saliency maps (SM), and (d, h) integrated gradients (IG). The preferable directions for changing explainability measures are marked by arrows.}
  \label{fig:fig3}
\end{figure}

The faithfulness of XAI techniques is determined via an erasure procedure\cite{li2016understanding}, which comprises removing some tokens and identifying changes in model confidence. We estimate two faithfulness measures. Specifically, comprehensiveness as an evaluation metric stands for model degradation caused by eliminating the most influential tokens; a larger value is better. On the other hand, sufficiency hinges on model stability if only influential tokens are taken into account; a smaller value is better. The tokens are excluded within a ranking produced by the explainer being analyzed. A more detailed description of both metrics is provided in the \hyperref[sec:explanation]{Methods} section. In the following explainability analysis, we limit our attention to the MatBERT model used in band gap classification (Figure \ref{fig:fig3}). With comprehensiveness, most explainers result in a bimodal distribution with peaks at zero and one. Therefore, there are two distinct groups of test examples differing in the ability of the XAI methods to extract meaningful rationales behind model reasoning. Entities with the score close to one are almost comprehensively interpreted within the corresponding explainer; the opposite is true for the second group of materials descriptions. LIME shows the largest proportion of test examples with high comprehensiveness (>0.9): 49\%. SHAP has the second-highest ratio of 25\%, whereas SM and IG yield only 2\% and 8\%, respectively. In the case of sufficiency, the score distributions of all explainers have the main peak located at zero. Entities with a preferable score value (<0.1) hold a share of 97\% (LIME), 95\% (SHAP), 86\% (SM), and 69\% (IG). Low sufficiency means that only a tiny set of tokens actually affects model output. To sum up, LIME and SHAP provide rationales that are (often) comprehensible and (nearly always) sufficient to explain how a classifier in question makes a decision.

\begin{figure}[ht]
  \centering
  \includegraphics[width=15.5cm]{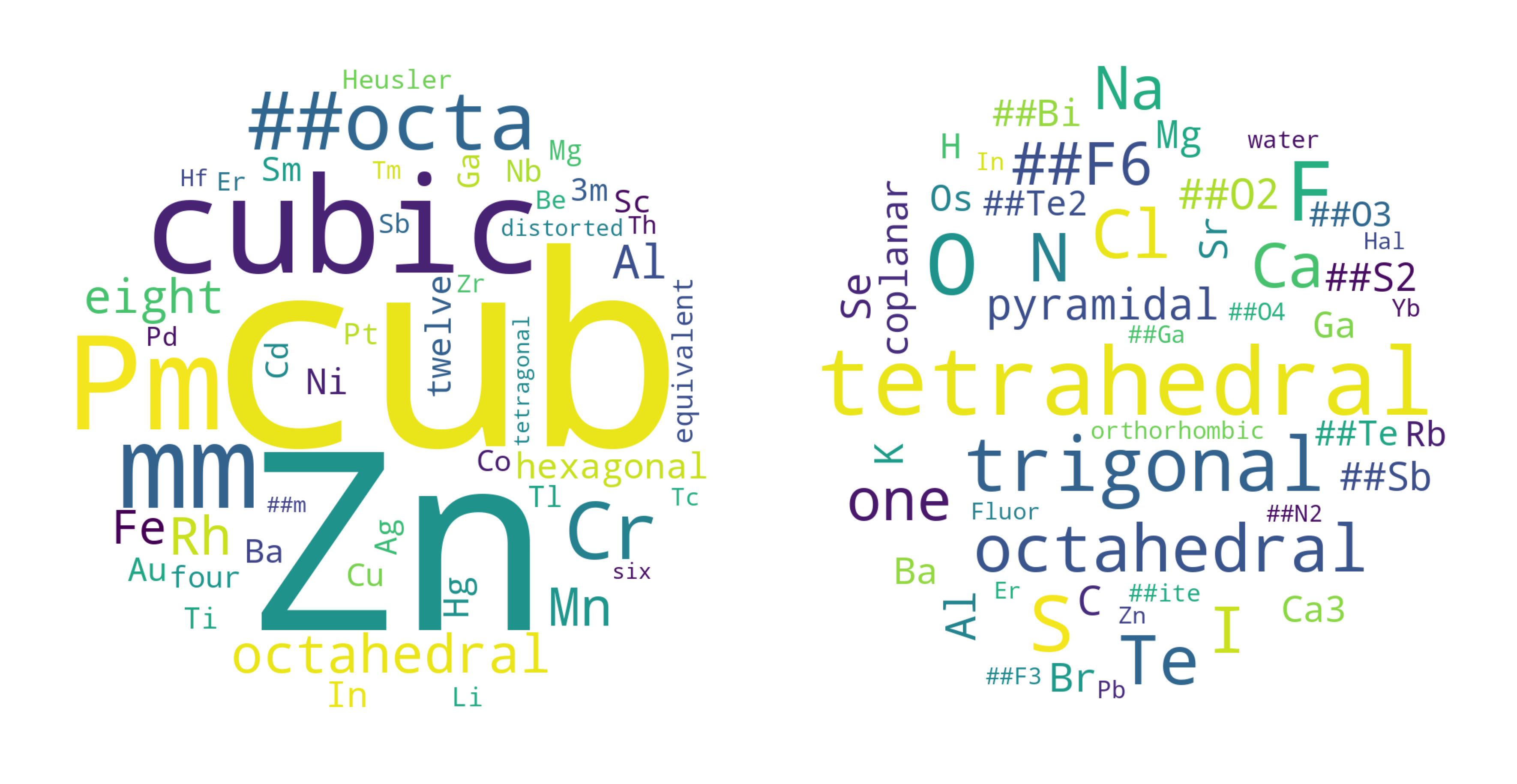}
  \caption{Most influential tokens in band gap classification provided by the \emph{post hoc} local explainer. Word clouds contain tokens that have the greatest impact on the MatBERT model decision to classify a material as a metal (left panel) or a nonmetal (right). The font size reflects the amplitude of the influence, whereas color differentiation helps to distinguish adjacent tokens.}
  \label{fig:fig4}
\end{figure}

The tokens highly ranked by the explainer with confirmed faithfulness may be helpful for explaining how a language model processes human-readable materials descriptions. We select the top 5\% of nonunique tokens within the ranking given by LIME. Two classes (metals vs nonmetals) are considered separately. Fifty most numerous unique tokens in both cases are finally examined (Figure \ref{fig:fig4}); the corresponding data for other classification tasks are presented in Figures S1–S4. The visualized words can be formally categorized into two groups. Namely, the former consists of chemical-element symbols and associated subtokens (\texttt{[\#\#Bi]}, \texttt{[\#\#Ga]}, \texttt{[\#\#Sb]}, and \texttt{[\#\#Te]}), while the latter contains the tokens accompanied with crystal structure. Although the MatBERT tokenizer was prepared on the in-domain corpus, it still cannot properly handle a minor part of chemical formulas. For instance, Rb$_{2}$IrF$_{6}$Hg decomposes into the following string of subtokens: \texttt{[Rb]}\texttt{[\#\#2]}\texttt{[\#\#Ir]}\texttt{[\#\#F6]}\texttt{[\#\#Hg]}. Therefore, the high priority of the \texttt{[\#\#F6]} subtoken may indicate not only an influence of fluorine on a target property but also the importance of specific stoichiometry. The same conclusion is also true for similar subtokens depicted in Figure \ref{fig:fig4}: \texttt{[\#\#F3]}, \texttt{[\#\#O3]}, \texttt{[\#\#Te2]}, and others. Unambiguous identification of stoichiometry’s impact would require further developments in tokenization of chemical formulas. The MatBERT tokenizer does not ideally parse space group symbols either. For this reason, tokens \texttt{[Pm]} and \texttt{[3m]} are present due to incorrect processing of some space groups that inherit inversion symmetry: $R3\overbar{m}$, $Pm3\overbar{m}$, $Fm3\overbar{m}$, and others. Token \texttt{[mm]} originates from splitting of space group symbols by a slash, e.g., $P4/mmm$ is split into \texttt{[P4]}\texttt{[/]}\texttt{[mm]}\texttt{[\#\#m]}. Here again, we cannot differentiate the importance of a specific space group and that of its inherent symmetry elements. Next, numbers as tokens (\texttt{[one]}, \texttt{[four]}, \texttt{[six]}, \texttt{[eight]}, and \texttt{[twelve]}) are related to the number of nearest neighbors of a described site. Then, the set of tokens, including \texttt{[coplanar]}, \texttt{[cubic]}, \texttt{[cub]}\texttt{[\#\#octa]}\texttt{[\#\#hedral]}, \texttt{[tetrahedral]}, \texttt{[octahedral]}, \texttt{[pyramidal]}, \texttt{[trigonal]}, \texttt{[water]}\texttt{[-]}\texttt{[like]}, and \texttt{[hexagonal]}, serves to describe a coordination environment. It partially overlaps with another set, which contains crystal systems: \texttt{[cubic]}, \texttt{[trigonal]}, \texttt{[hexagonal]}, \texttt{[tetragonal]}, and \texttt{[orthorhombic]}. Tokens \texttt{[Fluor]}\texttt{[\#\#ite]}, \texttt{[Hal]}\texttt{[\#\#ite]}, and \texttt{[Heusler]} refer to the eponymous structural types. Finally, tokens \texttt{[distorted]} and \texttt{[equivalent]} help to characterize local (dis)order.

\begin{figure}[ht]
  \centering
  \includegraphics[width=9cm]{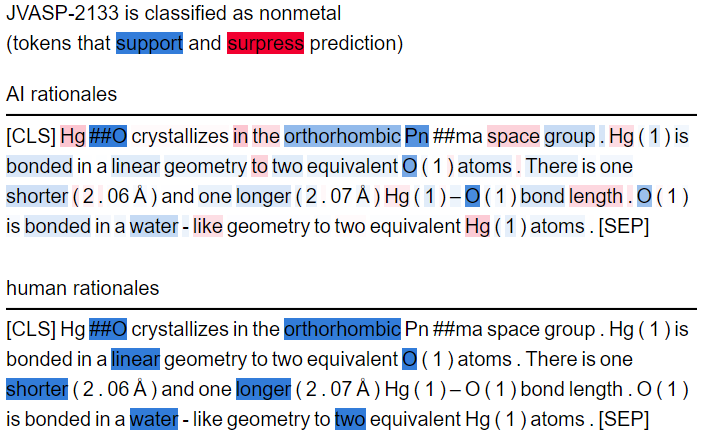}
  \caption{Token importance levels determined by a \emph{post hoc} local explainer and a domain expert. Distinct words are colored in accordance with their impact on model output; color intensity denotes the amplitude, whereas the color (red vs blue) means a contribution to the predicted class (negative vs positive).}
  \label{fig:fig5}
\end{figure}

Word clouds displayed in Figure \ref{fig:fig4} provide a bird’s-eye view of most general relations between a material’s features and its electronic structure (the presence of a band gap). MatBERT can reveal well-known patterns, such as the abundance of tetrahedral structures among semiconductors (nonmetals in our terminology) and dominance of intermetallic compounds as metals. To gain more insights into the plausibility of the presented language-centric approach, we directly compare most influential tokens extracted by a faithful explainer (LIME) with rationales proposed by a domain expert. It should be mentioned that the coauthor who highlighted meaningful tokens did not participate in ML model reasoning analysis. In this way, we sought to avoid a bias in human decision-making. The domain expert operates under a discrete regime assigning one of two scores to each token (Figure \ref{fig:fig5}): insignificant (0) or important (1). Then, his/her rationales are matched with continuous importance measures identified by the XAI method. Two plausibility metrics described in the \hyperref[sec:explanation]{Methods} section are calculated for a subset of test examples (287 entities, band gap classification): the token level F1 score and area under the precision–recall curve (AUPRC), which are equal to 0.33 and 0.32, respectively. Due to the absence of relevant XAI studies in the materials science field, we have to compare the obtained values with the values available in other scientific fields. In the Evaluating Rationales And Simple English Reasoning\cite{deyoung2019eraser} (ERASER) benchmark study, seven datasets covering diverse document types (from reports of clinical trials to movie reviews) are analyzed to quantify interpretability of several language models. Best-performing models have a token level F1 score and AUPRC in the range of 0.134–0.812 and 0.244–0.606, respectively. Thus, the consistence of our language model within domain expert reasoning is comparable to the previously obtained ones. We hope the present study will stimulate the creation of interpretability-aware benchmarks, resulting in an understanding of how one can reach high interpretability of ML algorithms in materials research. To facilitate such efforts, open access to the first expert-annotated corpus of materials descriptions for band gap predictions is provided.

Referring to a recent perspective\cite{krenn2022scientific}, the presented language-centric XAI framework interacts with two of three dimensions of AI-assisted scientific understanding. On the one hand, scientists are supposed to generalize insights gained from a “computational microscope” without the need for complete computation. New data representations with inherent transparency can promote advances in the field. Indeed, crystal structure descriptions placed in the context of supervised ML help us to reduce model reasoning to the concepts that are well known to materials scientists. On the other hand, the suggested approach falls into the second category of AI contributions. We explicate internal machinery of language models within XAI techniques, and this approach will enable researchers to obtain unexpected results by inspecting originally black-boxed algorithms in the future. In the aforementioned dimensions, AI serves as a consultant for human scientists. On the contrary, the algorithms belonging to the third dimension\cite{krenn2022scientific} are thought to be independent agents of understanding capable of translating their vision. So far, there are no algorithms that undoubtedly fall into this category. Nonetheless, taking into account recent advances in large language models such as ChatGPT\cite{van2023chatgpt}, ultra-strong XAI approaches\cite{muggleton2018ultra} have a bright future.

\section{Conclusion}
\label{sec:conclusion}
To sum up, here we present a language-centric framework aimed at accurately predicting materials properties and at providing clear explanations of the corresponding rationales simultaneously. State-of-the-art performance is grounded in the incorporation of domain knowledge into advanced transformer models through pretraining on a large corpus of papers. On the other hand, human-readable text-based descriptions as materials representation allow to compare model reasoning and expert decisions directly. The proposed approach offers an alternative to opaque ML models, which are omnipresent in materials informatics at present. At the concept level, our intention was to dispel the popular belief that AI techniques are black boxes unable to stimulate new insights into structure–property relationships.

\section{Methods}
\label{sec:methods}

\subsection{Datasets}
The JARVIS-DFT dataset, a part of JARVIS\cite{choudhary2020joint}, was the main source of data in the study. The vdW-DF-OptB88 van der Waals functional\cite{klimevs2009chemical} was used to calculate most materials properties; the Tran-Blaha modified Becke-Johnson functional\cite{tran2009accurate} was selected to reproduce better band gaps\cite{choudhary2018computational} and frequency-dependent dielectric functions and hence SLME\cite{choudhary2019accelerated}. The following classification tasks were analyzed: energy above the convex hull (55350 entities, a threshold of 0.1 eV/atom), the magnetic moment (52205, 0.05 $\mu$B), a band gap (18167, 0.01 eV), SLME (9063, 10\%), and topological spin-orbit spillage (11376, 0.1).

\subsection{Model training}
k-Fold cross-validation (10-fold) was performed to estimate model performance. In the case of attention-based neural networks trained on composition and graph neural networks, one-ninth of training data was retained as a validation set for early stopping, thereby ensuring a training-validation-test ratio of 80:10:10. The data from the validation set were not used in the training of all other models. We employed identical data splits for each prediction algorithm, considering a specific dataset. The analyzed classification metrics (accuracy, the F1 score, and MCC) were averaged over cross-validation subsets, while standard deviation was regarded as a measure of prediction uncertainty. All the deep learning models were developed within the PyTorch framework\cite{paszke2019pytorch}.

\emph{Random Forests on Force-Field Inspired Descriptors}. We trained random forest\cite{breiman2001random} on CFIDs\cite{choudhary2018machine}, including chemical, cell size, radial charge, and distribution (radial, angular, dihedral, and nearest-neighbor) features. The scikit-learn implementation\cite{pedregosa2011scikit} of the algorithm was chosen with default hyperparameter values. CFIDs were extracted using the JARVIS-Tools library\cite{choudhary2020joint}.

\emph{Attention-based Neural Networks on Composition}. A neural network referred to as the Roost framework\cite{goodall2020predicting} was trained on materials compositions represented as dense weighted graphs between elements. Node representations were updated through message-passage by the soft-attention mechanism. Fixed-length materials representations generated via a soft-attention pooling operation were then passed as input to a feed-forward network that finally generated an endpoint value. The AdamW optimizer\cite{loshchilov2017decoupled} with parameters ${\beta}_{1}$ = 0.9, ${\beta}_{2}$ = 0.999, and a learning rate of 10$^{-3}$ was employed. The model was trained for 1000 epochs with an early stopping at 100. The batch size was set to 128.

\emph{Graph Neural Networks}. We trained ALIGNN\cite{choudhary2021atomistic}, which was intended to capture many-body interactions explicitly. The ALIGNN architecture performed a series of edge-gated graph convolutions on an atomistic and corresponding line graph. The resulting atom representations were reduced by an average pooling operation and transferred to a fully connected network to predict a target property. The AdamW optimizer\cite{loshchilov2017decoupled} with parameters ${\beta}_{1}$ = 0.9, ${\beta}_{2}$ = 0.999, and a learning rate of 10$^{-3}$ was employed. The model was trained for 1000 epochs with an early stopping at 100. The batch size was set to 64. The original ALIGNN implementation was used that heavily relies on the Deep Graph Library\cite{wang2019deep}.

\emph{Transformer Language Models}. The human-readable descriptions of crystal structure were generated by means of the Robocrystallographer library\cite{ganose2019robocrystallographer}. The text information on the local (coordination number and geometry), semilocal (polyhedral connectivity and tilts angles), and global (mineral type and crystal symmetry) environments was represented as a sequence of tokens. The BERT\cite{devlin2018bert} model was chosen as a basic architecture. Taking into account weights’ initialization and tokenization procedures, three models were trained for each downstream task: a randomly initialized BERT model using the original tokenizer, a randomly initialized BERT model using the MatBERT tokenizer, and the MatBERT model\cite{trewartha2022quantifying} using the MatBERT tokenizer. Both case-sensitive tokenizers were based on the WordPiece algorithm\cite{wu2016google}. The AdamW optimizer\cite{loshchilov2017decoupled} with parameters ${\beta}_{1}$ = 0.9, ${\beta}_{2}$ = 0.999, and a learning rate of 3 $\times$ 10$^{-4}$ was utilized. The model was trained over 10 epochs with a batch size of 16. The HuggingFace  Transformers library\cite{wolf2020transformers} was extensively used to assess pretrained models and to fine-tune them.

\subsection{Model explanation}
\label{sec:explanation}
\emph{Explanation algorithms}. We took advantage of four XAI techniques. First, SMs\cite{simonyan2013deep} of the predicted class were generated. The method was originally formulated for image-specific class saliency visualization; here elements of SM, i.e., feature importance levels, were extracted at the token level as derivatives of predicted class probability with respect to the corresponding token embedding. Second, IGs\cite{sundararajan2017axiomatic} were defined as path integrals of the gradients along the straight-line path from the baseline (the padding token) to the considered token. Both SM and IG were calculated using the Captum package\cite{kokhlikyan2020captum}. Third, the LIME\cite{ribeiro2016should} approach was implemented to obtain token level importance scores. The predictors in question were approximated by a transparent algorithm: Ridge regression. Then, the surrogate model was optimized in such a way as to ensure both interpretability and local fidelity. We employed the original implementation of the algorithm for this purpose. Fourth, Shapley values\cite{shapley1953value} from game theory were assigned in order to quantify tokens’ contributions to the model outcome. To be precise, the extended version of Shapley values, aka Owen values\cite{owen1977values}, was computed to capture preferable input feature coalitions. Partition masking, as implemented in the SHAP package\cite{lundberg2017unified}, was applied for this purpose.

\emph{Evaluation metrics}. Faithfulness of ML predictors was assessed via two metrics\cite{deyoung2019eraser}. Starting with original sequence of tokens ${x}_{i}$, we constructed its contrast example by removing subset of tokens ${r}_{i}$. Comprehensiveness is defined as a difference between probability assigned by model $m$ to initial sequence of tokens ${x}_{i}$ and probability derived by the same algorithm from sequence with removed rationales ${x}_{i}\backslash{r}_{i}$:

\begin{equation}
  comprehensiveness = m({x}_{i})_{j} - m({x}_{i}\backslash{r}_{i})_{j}
\end{equation}

Sufficiency is oppositely defined as a difference between probability assigned by model $m$ to initial sequence of tokens ${x}_{i}$ and probability derived by the same algorithm from sequence of removed rationales ${r}_{i}$:

\begin{equation}
  sufficiency = m({x}_{i})_{j} - m({r}_{i})_{j}
\end{equation}

Both metrics were calculated for predicted class $j$, i.e., the class with the highest probability $m({x}_{i})_{j}$. The arbitrariness of the choice of subset ${r}_{i}$ was overcome as follows. We calculated faithfulness measures assuming subsets of rationales ${r}_{ik}$ that included $k$ percent of most important tokens identified by an explainer of interest. Then, an aggregate metric referred to as area over the perturbation curve\cite{deyoung2019eraser} (AOPC) was calculated as

\begin{equation}
  AOPC = \frac{1}{|\mathcal{B}| + 1} \left (\sum_{k=0}^{|\mathcal{B}|} m({x}_{i})_{j} - m({x}_{i}\backslash{r}_{ik})_{j} \right)
\end{equation}

Set of percentiles $\mathcal{B}$ is $\{ 10\%, 20\% , \ldots, 100\% \}$. Throughout the main text, by comprehensiveness and sufficiency we mean the corresponding AOPC values.

Plausibility was estimated for discrete and soft explanations\cite{deyoung2019eraser}. For each example, the subset of tokens selected by a domain expert was compared with the subset of rationales that included $k$ most influential tokens according to explainer ranking. The value of $k$ is set to the average rationale length proposed by a human (10). The corresponding token level F1 score was regarded as a plausibility measure. In addition, we estimated AUPRC to take into account tokens’ ranking.

All the explainability-related calculations were carried out within the ferret package\cite{attanasio2022ferret}.

\section{Data and Code Availability}
\label{sec:data}
The expert-annotated dataset for band gap classification is available at \url{https://zenodo.org/record/7750192}. The trained MatBERT model for band gap classification is available at \url{https://huggingface.co/korolewadim/matbert-bandgap}. All other data and code reported in this paper will be shared by the corresponding author upon request.

\bibliographystyle{unsrt}
\bibliography{references}

\end{document}